\documentclass[twocolumn,showpacs,preprintnumbers,amsmath,amssymb]{revtex4}

\usepackage{amsmath,amssymb,amsfonts,epsf}

\usepackage{graphicx}
\let\bm=\bibitem
\newcommand{\fft}[2]{{\frac{#1}{#2}}}
\newcommand{\ft}[2]{{\textstyle\frac{#1}{#2}}}
\def\nn{\nonumber}
\newcommand{\be}{\begin{equation}}
\newcommand{\ee}{\end{equation}}
\def\ft#1#2{{\textstyle{\frac{\scriptstyle #1}{\scriptstyle #2}}}}
\def\fft#1#2{\frac{#1}{#2}}
\def\td{\tilde}
\newcommand{\bea}{\begin{eqnarray}}
\newcommand{\eea}{\end{eqnarray}}

\begin{document}

\date{\today}
\vspace{2.7in}

\title{Solar Radiation Pressure and Deviations from Keplerian Orbits}
\author{Roman Ya. Kezerashvili and Justin F. V\'azquez-Poritz} \affiliation{\mbox{Physics Department, New York City College of Technology, the City University of New York, }\\
Brooklyn, NY 11201, USA\\}

\begin{abstract}

Newtonian gravity and general relativity give exactly the same expression for the period of an object in circular orbit around a static central mass. However, when the effects of the curvature of spacetime and solar radiation pressure are considered simultaneously for a solar sail propelled satellite, there is a deviation from Kepler's third law. It is shown that solar radiation pressure affects the period of this satellite in two ways: by effectively decreasing the solar mass, thereby increasing the period, and by enhancing the effects of other phenomena, potentially rendering some of them detectable. In particular, we consider deviations from Keplerian orbits due to spacetime curvature, frame dragging from the rotation of the sun, the oblateness of the sun, a possible net electric charge of the sun, and a very small positive cosmological constant.

\vspace{0.1cm}

\pacs{95.30.Sf, 91.10.Sp, 96.50.Ek, 96.60.Tf}

\end{abstract}

\maketitle {}

In the last decade, the observation and analysis of satellite motion has provided an abundance of data with which to test basic physical principles. Examples include the Pioneer anomaly, which is an unexplained acceleration of the Pioneer 10 and 11 spacecraft on escape trajectories from the outer solar system \cite{rk3,rk4}, and the flyby anomaly, for which the velocities of the Galileo, NEAR and Cassini spacecraft  are different from what is expected after Earth flybys \cite{AnderWilliams,Lammerzahal}. In fact, the difficulties of explaining these anomalies within the framework of standard physics became a motivation to speculate on the unlikely possibility that they originate from new physics. Missions have been proposed \cite{Lammerzahal,rk6,rk5} to further explore these anomalies, in order to better understand the laws of fundamental physics as they affect dynamics within the solar system. 

One of the most basic laws that describes motion in the solar system is Kepler's third law, which can be derived from Newton's law of gravitation and provides a relationship between the period $T$ and the orbital radius $r$ of an object orbiting the sun. Namely, $T^2\propto r^3$ with the proportionality constant given by $4\pi ^{2}/GM$, where $G$ is the gravitational constant and $M$ is the mass of the sun. Below, we discuss deviations from Keplerian orbits due to phenomena within conventional physics, some of which may be observed from the motion of solar sail propelled (SSP) satellites \cite{rk1,rk2}.  

In the Newtonian approximation, the sun is the source of a gravitational force on other masses. In the general relativistic framework, objects follow geodesics on the curved spacetime in the vicinity of the sun. At the same time, the sun is also a source of solar electromagnetic radiation, which produces an external force on objects via the solar radiation pressure (SRP). We will assume that the backreaction of the radiation on spacetime is negligible. Therefore, we can say that objects move in the {\it photo-gravitational} field of the Sun. 

The purpose of this Letter is to point out various sources of deviations from Keplerian orbits and discuss how the resulting change in period is enhanced by the SRP to the degree in which it may be observed for some cases. The phenomena discussed include the curvature of spacetime in the vicinity of the sun, described by the Schwarzschild metric, frame dragging due to the rotation of the sun, for which the curved spacetime is described approximately by the large-distance limit of the Kerr metric \cite{kerr}, the oblateness of the sun, the effect of a possible small net electric charge on  the sun, and a small positive cosmological constant. This list is by no means exhaustive. Some effects that we do not consider the gravitational field of planets, the magnetic field of the sun and the solar wind. While the reality is that all effects occur simultaneously, particular effects may be isolated by considering a variety of orbits. However, in order to get an approximate idea of the {\it relative} importance of some effects in the presence of the SRP, we will consider them individually for the simple scenario of circular orbits.

According to Maxwell's electromagnetic theory, electromagnetic waves carry the energy and linear momentum and the radiation pressure $P$ exerted on a surface due to momentum transport by photons from the sun is given by
\be
P=\fft{2\eta S}{c}\,,\qquad S=\frac{L_s}{4\pi r^{2}}\,,
\ee
where $S$ is the magnitude of the Poynting vector and the solar luminosity is $L_s=3.842\times 10^{26}$ W. Also, $0.5\le \eta\le 1$, where $\eta=0.5$ corresponds to the total absorption of photons by the satellite and $\eta=1$ corresponds to total reflection. The resulting force is $F=PA$, where $A$ is the surface area facing the sun. Thus, the acceleration due to this force can be expressed as
\be\label{kappa}
a=\fft{\kappa}{r^2}\,,\qquad \kappa\equiv \frac{\eta L_s}{2\pi c\sigma}\,,\qquad \sigma =\frac{m}{A}\,,
\ee
where $m$ is the mass of the object. In fact, the mass per area $\sigma$ is a key design parameter for solar sails \cite{rk1,rk11}. 

We will first consider the effect of the SRP on Keplerian orbits in the Newtonian approximation for gravity. The SRP force is repulsive and the gravitational force is attractive. Both forces act along the same line (we are restricting ourselves to the case in which the surface is directly facing the sun) and they both fall off as $1/r^2$. Therefore, the consideration of both forces leads to a modification of the effective mass of the sun in Kepler's third law. Namely, the mass of the sun, which is $M=1.99\times 10^{30}$ kg, can be effectively renormalized as $\td M\equiv M-\kappa/G$. The modified Kepler's third law can be expressed as
\be
T^2=\fft{4\pi^2}{G\td M}\ r^3\,.
\ee

As a first example, we will consider Mercury, whose average orbital radius is $r=5.79\times 10^{10}$ m,  which corresponds to a period of about $88$ days. Mercury has a mass of $3.30\times 10^{23}$ kg and a radius $r_M=2.44\times 10^6$ m. This yields $\sigma=1.76\times 10^{10}$ kg/m$^2$, where the effective area is $\pi r_M^2$. If we assume that no sunlight is reflected by Mercury, so that $\eta=0.5$, then the increase in period is on the order of $10^{-7}$ s which, as to be expected for any planet, is negligible. 

We will now consider a conventional satellite orbiting the sun at a distance of $r=1$ AU$\approx 1.50\times 10^{11}$ m, which corresponds to a period of one year. If the mass of the satellite is $1000$ kg and its area is $2$ m$^2$, then $\sigma=500$ kg/m$^2$. If we suppose that $\eta=0.75$, then the increase in period due to SRP is about $36$ s, which could be observed.

In the remainder of the Letter, we will consider an SSP satellite with the following specifications:
\bea\label{values}
r &=& 0.05\ \mbox{AU}\approx 7.48\times 10^9\ \mbox{m}\,,\nn\\
\eta &=& 0.85\,,\quad \sigma=1.31\times 10^{-3}\ \mbox{kg/m}^2\,.
\eea
If the acceleration due to the SRP is ignored, then the corresponding orbital period would be about $4$ days. When the SRP is taken into account, the period is about $70$ days. 

We will now consider the simultaneous effects of the SRP and curved spacetime in the vicinity of the sun. We assume that the backreaction of the electromagnetic radiation on the background geometry is negligible so that it acts on the SSP satellite only via the SRP. The static exterior spacetime of the sun is described by the Schwarzschild metric, which is given by
\be\label{form}
ds^2=-f c^2dt^2+f^{-1} dr^2+r^2 d\Omega^2\,,
\ee
where
\be\label{f}
f=1-\fft{2GM}{c^2r}\,,\qquad d\Omega^2=d\theta^2+\sin^2\theta\ d\phi^2\,.
\ee

Spherical symmetry allows us to orient the coordinate system so that the orbit is confined to the equatorial plane at $\theta=\pi/2$, and thus $p_{\theta}=0$. Since the metric is independent of time and the azimuthal direction $\phi$, the corresponding components $p_t$ and $p_{\phi}$ of the 4-momentum are conserved. We define the constants of motion $E\equiv -p_t/m$ and $L\equiv p_{\phi}/m$, where $m$ is the rest mass of the SSP satellite. Note that $E$ and $L$ are the energy and angular momentum of a unit mass, respectively. Thus,
\be
p^t =\fft{m E}{c^2f},\quad p^r=m\ \fft{dr}{d\tau},\quad p^{\theta}=0,\quad p^{\phi}=\fft{m}{r^2}L,
\ee
where $\tau$ is the proper time. In the absence of the SRP, $p^2=-m^2c^2$ yields
\be\label{reqn1}
\Big( \fft{dr}{d\tau}\Big)^2=\fft{E^2}{c^2}-\Big(c^2+\fft{L^2}{r^2}\Big) f\,.
\ee
Differentiation of (\ref{reqn1}) with respect to $\tau$ gives the radial component of the 4-acceleration
\be\label{firstar}
a^r=\fft{d^2r}{d\tau^2}+\fft{GM}{r^2}-\fft{L^2}{r^3}+\fft{3GM L^2}{c^2r^4}\,.
\ee

We will now turn on the SRP, so that
\be\label{ar}
a^r=\fft{\kappa}{r^2}\,,
\ee
where $\kappa$ is given in (\ref{kappa}). Note that even though the coordinate $r$ does not measure the proper distance, the surface area of a sphere is still given by $4\pi r^2$, which means that the magnitude of the Poynting vector as well as the acceleration are given by the same expressions as in the Newtonian approximation. Equating the expressions for $a^r$ given in (\ref{firstar}) and (\ref{ar}) and taking the first integral gives
\be\label{sailreqn}
\Big( \fft{dr}{d\tau}\Big)^2=\fft{E^2}{c^2}-c^2+\fft{2G\td M}{r}-\fft{L^2}{r^2}f\,.
\ee
Note that the SRP reduces the effective mass only in the term which is present for Newtonian gravity. 

For circular orbits,
\be\label{circular}
\fft{dr}{d\tau}=0\,,\qquad \fft{d^2r}{d\tau^2}=0\,.
\ee
This yields
\be
E^2 = c^4+\left( \fft{4GM-c^2r}{c^2r-3GM}\right) \fft{c^2G\td M}{r},\quad
L^2 = \fft{G\td M r^2}{r-3GM/c^2}.
\ee
Using $dt/d\tau=p^t/m$ and $d\phi/d\tau=p^{\phi}/m$, we find the orbital period $T$ to be given by the expression
\be\label{period}
T^2=\fft{4\pi^2r^3}{G\td M} \Big[ 1+\kappa\ \fft{c^2r-4GM}{(c^2r-2GM)^2}\Big]\,.
\ee
Keeping only the leading correction due to the curvature of spacetime gives
\be
T^2\approx\fft{4\pi^2r^3}{G\td M} \Big[ 1+\fft{\kappa}{c^2r}\Big]\,.
\ee
Notice that the simultaneous effects of the SRP and curved spacetime lead to a deviation from Kepler's third law, which is completely absent without the SRP. For the specification given in (\ref{values}), we find that this yields an increase in the period of about $0.6$ s.

The rotation of the sun causes frame dragging, which affects the trajectories of orbiting objects. 
The external spacetime of a slowly rotating body with mass $M$ and angular momentum $J$ is described approximately by the large-distance limit of the Kerr metric \cite{kerr}
\be
ds^2=-f c^2 dt^2-\fft{4GJ}{c^2r}\sin^2\theta\ dt d\phi+\fft{dr^2}{f}+r^2 d\Omega^2,
\ee
where $f$ is given by (\ref{f}). We do not use the full Kerr metric since it does not seem to describe the external spacetime of a rotating material body, because it does not smoothly fit onto metrics which describe the interior region occupied by physical matter. Since there are corrections to this metric in higher-order $J$ we will work up to only linear order in $J$, which suffices for the slowly rotating sun. 
Note that $J>0$ for a prograde orbit with respect to the sun, while $J<0$ for a retrograde orbit. 

Restricting ourselves to orbits that lie within the equatorial plane, we define the constants of motion $E\equiv -p_t/m$ and $L\equiv p_{\phi}/m$ \cite{carter,bardeen}. We have
\bea
p^t &=& g^{tt} p_t+g^{t\phi}p_{\phi}=\fft{mE}{c^2f}-\fft{2GmJL}{c^4fr^3}\,,\nn\\
p^{\phi} &=& g^{\phi\phi}p_{\phi}+g^{t\phi}p_t=\fft{mL}{r^2}+\fft{2GmJE}{c^4fr^3}\,,
\eea
and $p^r=m\ dr/d\tau$. In the absence of the SRP, $p^2=-c^2m^2$ yields
\be
\Big( \fft{dr}{d\tau}\Big)^2=\fft{E^2}{c^2}-\Big( c^2+\fft{L^2}{r^2}\Big) f-\fft{4GJEL}{c^4r^3}\,.
\ee
Differentiating this with respect to $\tau$ gives
\be\label{kerrar}
a^r= \fft{d^2r}{d\tau^2}+\fft{GM}{r^2}-\fft{L^2}{r^3}+\fft{3G(c^2ML^2-2JEL)}{c^4r^4}\,.
\ee
Turning on the SRP, the equation of motion is given by (\ref{ar}), which is not altered by linear terms in $J$. Equating the expressions for $a^r$ given in (\ref{ar}) and (\ref{kerrar}) and taking the first integral gives
\be\label{reqnkerr}
\Big( \fft{dr}{d\tau}\Big)^2= \fft{E^2}{c^2}-c^2+\fft{2G\td M}{r}-\fft{L^2}{r^2}f -\fft{4GJEL}{c^4r^3}\,.
\ee
Applying the conditions for a circular orbit given in (\ref{circular}),
\bea
E &=& c \sqrt{\fft{X}{r(c^2r-3GM)}} - \fft{cJ}{r} \sqrt{\fft{G^3\td M}{(c^2r-3GM)^3}},\\
L &=& cr \sqrt{\fft{G\td M}{c^2r-3GM}} - \fft{3GJ}{c} \sqrt{\fft{X}{r(c^2r-3GM)^3}},\nn
\eea
where $X \equiv c^4r^2-c^2G(3M+\td M)r+4G^2M\td M$. Using $dt/d\tau=p^t/m$ and $d\phi/d\tau=p^{\phi}/m$, we find the orbital period to be
\bea
T^2 &\approx& \fft{4\pi^2 r^3}{G\td M} \Big[ 1+\kappa \fft{(c^2r-4GM)}{(c^2r-2GM)^2}\Big]\\ &\times& \Big[ 1+\fft{2\sqrt{G} J\Big( 1+\fft{4\kappa (c^2r-GM)}{(c^2r-2GM)^2}\Big)}{c^2\sqrt{\td M} r^{3/2} \sqrt{1+\fft{\kappa (c^2r-4GM)}{(c^2r-2GM)^2}}}\Big]+{\cal O}(J^2)\,.\nn
\eea
Keep only the leading contributions due to spacetime curvature and frame dragging gives
\be
T^2 \approx \fft{4\pi^2 r^3}{G\td M} \Big[ 1+\fft{\kappa}{c^2r}\Big] \Big[ 1+\fft{2\sqrt{G} J}{c^2\sqrt{\td M} r^{3/2}}\Big]\,.
\ee
The speed of the outer layer of the sun at its equator is $v\approx 2000$ m/s at the equatorial radius $R\approx 7\times 10^8$ m. If we assume that the core of the sun rotates with the same angular speed, then $J=\ft25 Mv R\approx 10^{42}$ kg m$^2$/s. Without the effects of the SRP, frame dragging leads to an increase (decrease) in the period of only $4\times 10^{-5}$ s for a prograde (retrograde) orbit. With the SRP and the specifications in (\ref{values}), frame dragging leads to a change in the period of about $0.01$ s. 

We will now consider the effect of the oblateness of the sun. Working in Newtonian gravity, the external gravitational potential of an oblate spheroid is given by \cite{oblateness}
\be
V=-\fft{G\td M}{r}+\fft{GM}{r}\sum_{n=2}^{\infty} J_n \Big( \fft{R}{r}\Big)^n P_n (\cos\theta)\,,
\ee
where $J_n$ are the multipole mass moments, $R$ is the equatorial radius of the sun and $P_n$ are the Legendre polynomials. Note that the effective mass in the first term is renormalized by the solar radiation pressure, whereas the multipole mass moments are not affected. We will consider only the case for which the orbit of the SSP satellite is confined to the equatorial plane. Then the acceleration is purely in the radial direction, given by
\be
a^r=-\fft{G\td M}{r^2}\left(1-\fft{3GMJ_2 R^2}{2\td Mr^2}+\fft{15MJ_4 R^4}{8\td M r^4}+\cdots\right)\,,
\ee
This leads to the relation
\be
T^2\approx\fft{4\pi^2r^3}{G\td M} \Big[ 1+\fft{3MJ_2R^2}{2\td Mr^2}\Big]\,,
\ee
where we consider the first sub-leading term due to the oblateness. The oblateness of the sun has recently been measured with unprecedented precision to be as much as $J_2\approx 9\times 10^{-6}$ during active phases of the solar cycle \cite{oblatesun}. Without the SRP, the oblateness of the sun increases the period by about $0.02$ s. With the SRP and the specifications given in (\ref{values}), the sun's oblateness results in an increase in the period of about $105$ s. It can be shown that the next mass moment $J_4$ increases the period by only about $0.0006$ s.

We will now consider the effect of a small amount of net charge $Q$. It has been suggested that the sun has a net charge of  up to $Q\approx 77$ C \cite{solarcharge}. The spacetime is described by the Reissner-Nordstr\"om metric, which has the form (\ref{form}), where the function $f$ is now given by
\be
f=1-\fft{2GM}{c^2r}+\fft{Gk Q^2}{c^4r^2}\,,
\ee
and $k$ is the Coulomb constant. Taking into account the electric force, we find that
\be
T^2\approx\fft{4\pi^2r^3}{G\td M} \Big[ 1+\fft{\kappa}{c^2r}+\fft{kqQ}{Gm\td M}+\fft{kQ^2}{c^2\td M r}+\fft{k^2q^2 Q^2}{G^2m^2 \td M^2}+\cdots \Big],
\ee
where $q$ is the charge of the solar sail. Note the $Q^2$ term in the period, due to the backreaction from the sun's charge on the geometry, is present even for a neutral SSP satellite. However, this term increases the period by an amount of only $10^{-35}$ s ($10^{-38}$ s without the SRP), which reflects the fact that the backreaction of the charge on the geometry is negligible, as to be expected. 

Primarily due to the photoelectric effect, but also the Compton effect and electron-positron pair production, a solar sail made from Beryllium, for example, will equilibrate to a charge per area 
of $0.065$ C/m$^2$ \cite{rk9,rk10}. If the SSP satellite has a mass of $1000$ kg, this gives a charge $q=5\times 10^4$ C and an increase in period of about $230$ s ($0.05$ s without the SRP). Thus, due to the SRP, even 
a small net charge $Q$ could yield a measurable increase in the period. However, the challenge would be to isolate the effects of the sun's highly variable magnetic field.

Supernovae observations suggest that our universe might have a very small positive cosmological constant  \cite{cosm1,cosm2}. Since the SRP enhances a variety of small effects, one could ask how much the effects of a cosmological constant are enhanced. In the presence of a cosmological constant $\Lambda$, the spacetime in the vicinity of the sun is described by a metric of the form (\ref{form}), where the function $f$ is now given by
\be
f=1-\fft{2GM}{c^2r}-\fft{\Lambda r^2}{3}\,.
\ee
Considering the effect of $\Lambda$ for $r\gg GM/c^2$, we get
\be
T^2\approx\fft{4\pi^2r^3}{G\td M} \Big[ 1+\fft{c^2\Lambda}{3G\td M}r^3\Big]\,.
\ee
For $\Lambda\approx 10^{-52}$ m$^{-2}$, this leads to an insignificant increase in period of $10^{-17}$ s ($10^{-21}$ s without the SRP). While one might be able specify values of $\sigma$ and $r$ such that the correction factor is actually significant, the period would then be too large to make this a feasible test for the presence of a cosmological constant.

Below we summarize our results for the change in period $\Delta T$ of an SSP satellite with specifications (\ref{values}).
\vspace{.4cm}
\begin{center}
\begin{tabular}{|c|c|c|}
\hline Phenomenon & $\Delta T$ without SRP & $\Delta T$ with SRP \\
\hline\hline Spacetime Curvature & $0$ s & $0.6$ s \\
\hline Frame Dragging & $\pm 4\times 10^{-5}$ s & $\pm 0.01$ s \\
\hline Oblateness of Sun & $0.02$ s & $105$ s \\
\hline Net Charge of Sun & $0.05$ s & $230$ s \\
\hline Cosmological Const. & $10^{-21}$ s & $10^{-17}$ s \\
\hline
\end{tabular}
\end{center}
\vspace{.4cm}
The effects of some phenomena are increased by several orders of magnitude due to the SRP. While this could be sufficient to observe certain phenomena, the SRP would also enhance additional effects that we have not discussed, and a variety of orbits would be needed in order to isolate particular effects. Observations of the above deviations in the period would provide an interesting confirmation of these phenomena.


\begin{thebibliography}{99}

\bm{rk3} J. D. Anderson, P. A. Laing, E. L. Lau, A. S. Liu, M. M. Nieto and S. G. Turyshev, Phys. Rev. Lett. {\bf 81}, 2858-2861, 1998.

\bm{rk4} J. D. Anderson, et al., Phys. Rev. D {\bf 65}, 082004/1-50, 2002.

\bm{AnderWilliams}J. D. Anderson and J. G. Williams, Class. Quantum Grav. {\bf18}, 2447-2458, 2001. 

\bm{Lammerzahal} C. L\"ammerzahl, O. Preuss and H. Dittus, arXiv: gr-qc/ 0604052, 2006.

\bm{rk6} D. Izzo and A. Rathke, Journal of Spacecraft and Rockets {\bf 43}, 806-821, 2006. 

\bm{rk5} H. Dittus, S. G. Turyshev, C. L\"ammerzahl, et. al., arXiv: gr-qc/0506139, 2005.

\bm{rk1} C. R. McInnes, Solar Sailing. Technology, Dynamics and Mission
Applications, Springer, Praxis Publishing, 296 p., 1998.

\bm{rk2} G. L. Matloff, Deep space probes. To outer solar system and
beyond, 2nd. ed., Springer-Praxis, Chichester, UK, 242 p., 2005.

\bm{kerr} R. P. Kerr, Phys. Rev. Lett. {\bf 11}, 237-238, 1963.

\bm{rk11} R. Ya. Kezerashvili, JBIS, {\bf61}, 430-439, 2008; Acta Astronautica, in press (2009), doi: 10.1016/j.actaastro.2009.01.062.

\bm{carter} B. Carter, Phys. Rev. {\bf 174}, 1559-1571, 1968.

\bm{bardeen} J. M. Bardeen, W. H. Press and S. A. Teukolsky, Astrophys. J {\bf 178}, 347-369, 1972.

\bibitem{oblateness} L. D. Landau and E. M. Lifshitz, Teorya Polya,\\ FIZMATGIZ, Moscow 1939.

\bibitem{oblatesun} M. D. Fivian, H. S. Hudson, R. P. Lin and H. J. Zahid, Science {\bf 322}, 5901, 560-562, 2008.

\bibitem{solarcharge} L. Neslu\v{s}an, Astronomy and Astrophys. {\bf 372}, 913-915, 2001.

\bm{rk9} R. Ya. Kezerashvili and G. L. Matloff, JBIS, {\bf60}, 169-179, 2007.

\bm{rk10} R. Ya. Kezerashvili and G. L. Matloff, JBIS, {\bf61}, 47-57, 2008.

\bibitem{cosm1} S. Perlmutter {\it et al}. [Supernova Cosmology Project Collaboration], Astrophys. J. {\bf 483}, 565-581, 1997.

\bibitem{cosm2} A. G. Riess {\it et al}. [Supernova Search Team Collaboration], Astron. J. {\bf 116}, 1009-1038, 1998.

\end{thebibliography}
\end{document}